\documentclass[letter,bibyear]{aa} 

\usepackage{graphicx}
\usepackage{txfonts}
\begin{document}

   \title{On the location of the supermassive black hole in CTA\,102}


   \author{Christian M.\ Fromm
          \inst{1}
          \and
	  Manel Perucho
	  \inst{2,3}
	  \and
	  Eduardo Ros
	  \inst{1,2,3}
	  \and
	  Tuomas Savolainen
     	  \inst{1,4}
	  \and
	  J.\ Anton Zensus
	  \inst{1}
          }

   \institute{
Max-Planck-Insitut f\"ur Radioastronomie, Auf dem H\"ugel 69, 53121 Bonn, Germany\\
\email{cfromm, ros, tukasa, azensus@mpifr.de}
\and
Departament d'Astronomia i Astrof\'{\i}sica, Universitat de Val\`encia, C/ Dr. Moliner 50, 46100 Burjassot (Val\`encia), Spain\\
\email{Manel.Perucho@uv.es}
\and
Observatori Astron\`omic, Parc Cient\'{\i}fic, Universitat de Val\`encia, C/ Catedr\`atic Jos\'e Beltr\'an 2, E-46980 Paterna (Val\`encia), Spain 
\and
Aalto University Mets\"ahovi Radio Observatory, Mets\"ahovintie 114, FIN-02540 Kylm\"al\"a, Finland
}

   \date{
Draft 1.0: \today
}

 
  \abstract
{Relativistic jets in active galactic nuclei represent one of the most powerful
phenomena in the Universe.
They form in the surroundings of the supermassive black holes as a by-product of accretion onto the central black hole in active galaxies.
The flow in the jets propagates at velocities close to the speed of light.
The distance between the first part of the jet that is visible in radio images (core) and the black hole is still
a matter of debate.}  
{Only very-long-baseline interferometry observations resolve the innermost
compact regions of the radio jet.  Those can access the jet base, and combining
data at different wavelenghts, address the physical parameters of the 
outflow from its emission.}
{We have performed an accurate analysis of the frequency-dependent shift of the VLBI core location for a multi-wavelength set of
images of the blazar \object{CTA\,102} including data from 6\,cm down to 3\,mm.}
{The measure of the position of the central black hole, with mass $\sim10^{8.93}\,M_\odot$, in the blazar \object{CTA\,102} reveals a distance of $\sim 8\times10^4$ gravitational radii to the 86~GHz core, in agreement with similar measures obtained for other blazars and distant radio galaxies, and in contrast with recent results for the case of nearby radio galaxies, which show distances between the black hole and the radio core  that can be two orders of magnitude smaller.}
   {}


   \maketitle
%

\section{Introduction}

The blazar \object{CTA\,102} has a redshift $z=1.037$ (Schmidt \cite{sch65}; a luminosity distance of 6.7\,Gpc, and an image scale of {8.11}\,pc/milliarcsec)
and hosts a supermassive black hole (SMBH) of $10^{8.93}\,M_\odot$ (Zamaninasab et al.\ \cite{zam14}).
The source was one of the first to show strong variability in the radio after its discovery
in the 1960s.  Its radio morphology shows a strong core with
a southward jet on pc-scales, and two lobes in north-west and south-east directions
(e.g., Stanghellini et al.\ \cite{sta98}, Fromm et al.\ \cite{fro13a}).

The radio core of a jet is defined as its first observed surface at a given frequency
out of the black hole neighbourhood.
The position at which it appears at each frequency may be different if
the core corresponds to the first region that is optically thin to radiation at
this frequency, the lower frequencies appearing downstream of the higher ones,
as the jet flow becomes diluter and more transparent
(Marcaide \& Shapiro \cite{mar84}, Lobanov \cite{lob98}).
On the contrary, if the core corresponds to a strong
reconfinement shock, the core at the highest frequencies should
converge at the position of the shock.
A recent study (Hada et al.\ \cite{had11}, Marscher \cite{mar11}) 
on M\,87 suggested that
the core corresponds to the first optically thin surface and that the
43\,GHz surface is at 14 gravitational radii ($R_{\rm s}$) from the central black hole.
On the contrary, this distance has been suggested to be of $10^4 - 10^5$ gravitational radii in
the case of blazars
(Marscher et al.\ \cite{mar08,mar10}) and more distant radio galaxies than M87 such as 3C~111 and 3C~120 (Marscher et al. \cite{mar02}, 
Chatterjee et al. \cite{chat09,chat11}).

Fundamental differences between radio galaxies and blazar jets have
been invoked to explain the difference 
(Hada et al.\ \cite{had11}; Marscher et al.\ \cite{mar11}).
In blazars, the jet is observed at a very small angle to the line of sight, which favors
the observation of the fast flow due to the relativistic Doppler boosting of the
radiation (e.g., Zensus \cite{zen97}).
In radio galaxies, the
jet is typically observed at a larger angle, and the faster flow could be missed due to de-boosting of the radiation.
The observed jet in the two types of source could correspond to the slower,
outer layers of the jet, originated in the accretion disk surrounding the central
black hole in the case of radio galaxies, and to the faster spine of the jet generated
closer the the black hole, respectively.
Therefore, the physical properties of the observed radio core in the two types of
objects could be different.

In this letter, we use a 86~GHz VLBA observations made before the start of the flare as observed at 220~GHz (2005.6, Fromm et al. \cite{fro11}) together with observations down to 5~GHz to measure the distance from the black-hole to the radio-core at 86~GHz. This observation made during a quiescent state of the jet provides an opportunity to measure the core shift in a "clean" setting without the potential confusing effects of the flaring component. We expect changes in the opacity of the region during the flare, if it is followed by the propagation of a component, thus changing the measured distances following the core-shift method (Lobanov \cite{lob98}). 
Moreover, Fromm et al. (\cite{fro13b}) showed that the core-shift vectors on the plane of the sky presented a very irregular behaviour during the 2006 flare. The comparison of the steady state with these epochs would require a detailed understanding of the non-radial motions at the core, which is out of the scope of this paper. We refer the reader to recent works that show this behaviour, known as \emph{jet wobbling}, in other sources (Agudo et al. \cite{ag07,ag12}, Molina et al. \cite{mol14}). This letter is structured as follows: In Section~2 we present the relevant data for this work; in Section~3 we present our results; Section~4 is devoted to a brief discussion on possible effects that may have an influence on the result, and in Section~5 we summarize our work.


\section{VLBA observations and data analysis}

\begin{figure*}
\includegraphics[angle=90,width=0.99\textwidth,clip]{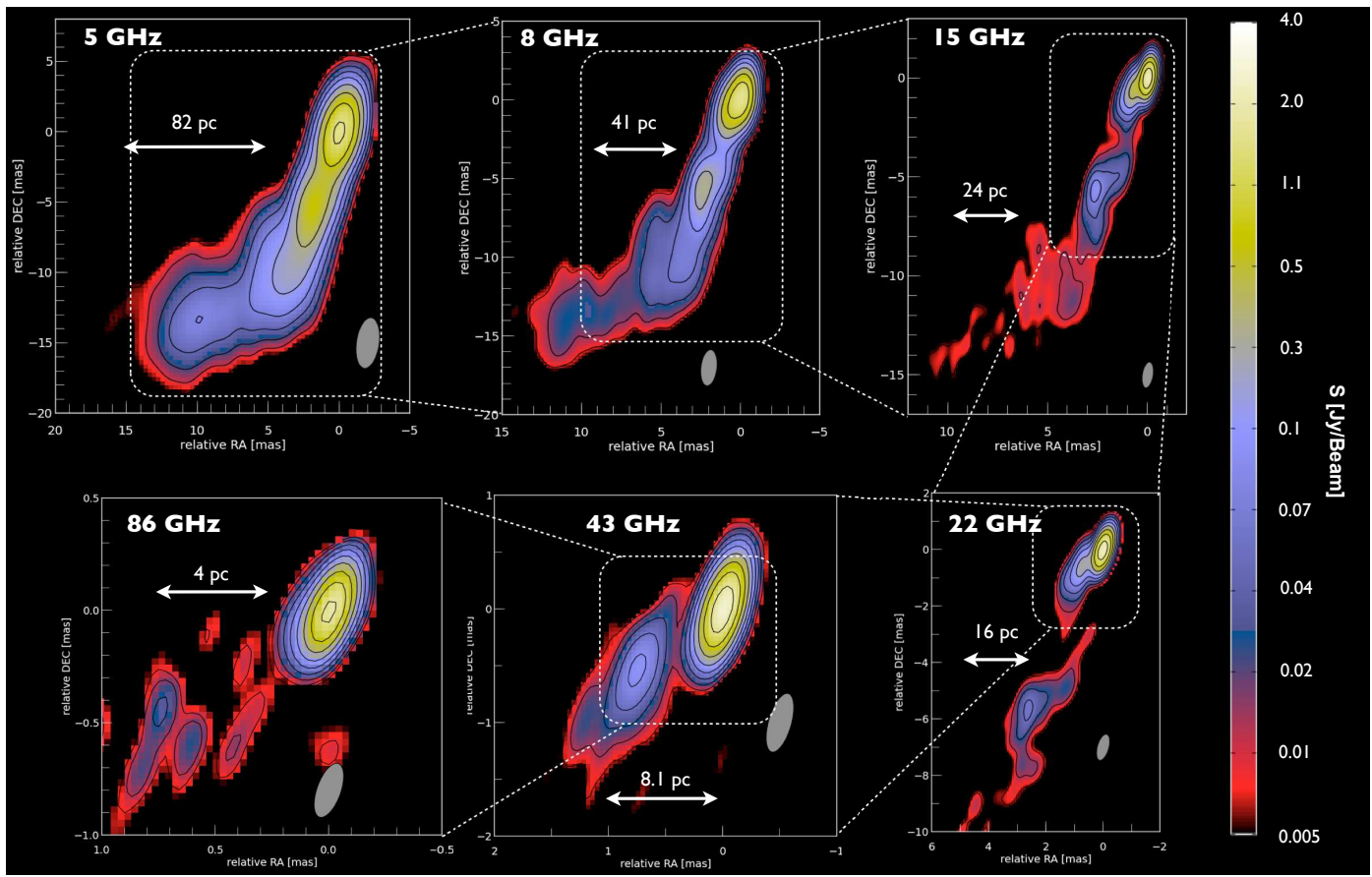}
\caption{\label{fig:coreshiftallfreq}
Parsec-scale radio images of the blazar \object{CTA\,102} as observed with
the VLBA on 2005-09-19. The interferometric beam is drawn at the bottom, right of
each image, as well as the linear scales. Contours are drawn on the false colour
image, being the lowest one at $0.005\,\mathrm{Jy/beam}$, and
spaced in factors of 2.
At the redshift of $z=1.037$ (luminosity distance of 6.9\,Gpc) 1\,mas
corresponds to a distance of 8.11\,pc and $8.4\times10^4$ gravitational radii for a
black hole mass of
$10^{10}$\,M$_\odot$.
}
\end{figure*}

We observed the jet in the blazar \object{CTA\,102} using the Very Long Baseline Array (VLBA)
at different frequencies, ranging from 5\,GHz to 86\,GHz. The results from these observations were 
presented in  Fromm et al. (\cite{fro11,fro13a,fro13b}), but for the case of the 86~GHz data, which is presented in Fig.~\ref{fig:coreshiftallfreq}.
Those observations cover two years around a major flare in the
source in 2006 (Fromm et al. \cite{fro11}). 
The 86~GHz epoch that we use here (2005.4) is the only one out of eight observations that yields high enough signal-to-noise ratio in the extended jet emission to allow alignment of the 86 GHz image with the lower frequencies.
In addition, the strong flares observed in many radio-sources are usually related with the later
detection of a region of enhanced emission that travels along the jet and can be
followed by fitting the interferometric data with a Gaussian function (usually
called \emph{component}) at each epoch.
These injected components can be related to an increase of the number of particles injected (Perucho et al. \cite{per08}),
which can affect the opacity in the region and change the relative position of the core
at different frequencies (Kovalev et al. \cite{kov08}).

 Figure~\ref{fig:coreshiftallfreq} shows the core region at all frequencies for epoch 2005.4. We aligned the images of the jet at different frequencies at each epoch using a cross-correlation method based on the optically thin jet regions (Croke \& Gabuzda \cite{cro08}; Fromm et al.\ \cite{fro13b}). This analysis revealed a two-dimensional shift of the core that can only be explained by non-axial (pattern or flow) motion of the emitting region (Fromm et al.\ \cite{fro13b}; see also, e.g., Agudo et al. \cite{ag07}, \cite{ag12}, Perucho et al. \cite{per12}).

\begin{table*}[t!]
\label{tab:coreshift}
\caption{Core shift values relative to reference (highest-observed) frequency for
different epochs}
\label{coreshiftstab1}
\centering
\resizebox{\hsize}{!}{
\begin{tabular}{@{}c c c c c c c c c c c c@{}}
\hline\hline
Epoch & $\nu_\mathrm{ref}$ &\multicolumn{2}{c}{$43\,\mathrm{GHz}$} &
\multicolumn{2}{c}{$22\,\mathrm{GHz}$} &  \multicolumn{2}{c}{$15\,
\mathrm{GHz}$} &  \multicolumn{2}{c}{$8\,\mathrm{GHz}$} &
\multicolumn{2}{c}{$5\,\mathrm{GHz}$} \\
& & r& PA &r&PA&r&PA& r& PA &r&PA\\
$[$yyyy-mm-dd] &[GHz]& [mas] & [$^\circ$] &[mas] & [$^\circ$] & [mas]
&[$^\circ$] & [mas] & [$^\circ$] &[mas] & [$^\circ$] \\
\hline
2005-05-19 &86& 0.03$\pm$0.04& 6 & 0.10$\pm$0.06 & 40 & 0.20$\pm$0.08 & 103 &0.32$\pm$0.11 & 74& 0.57$\pm$0.16 &90\\
\hline
\multicolumn{12}{l}{Notice, that the uncertainties presented here differ from the ones presented in Fromm et al. \cite{fro13b}. Here we used a more }\\
\multicolumn{12}{l}{conservative estimate for the uncertainties of the image alignment. The PAs are given in the definition of sky-plane}\\
\multicolumn{12}{l}{(counter-clockwise from North) not in the mathematical one as in Fromm et al. \cite{fro13b}.}
\end{tabular}
}
\end{table*}

\begin{figure}[t!]
\label{fig:totalcoreshift}
\includegraphics[width=0.99\columnwidth,clip]{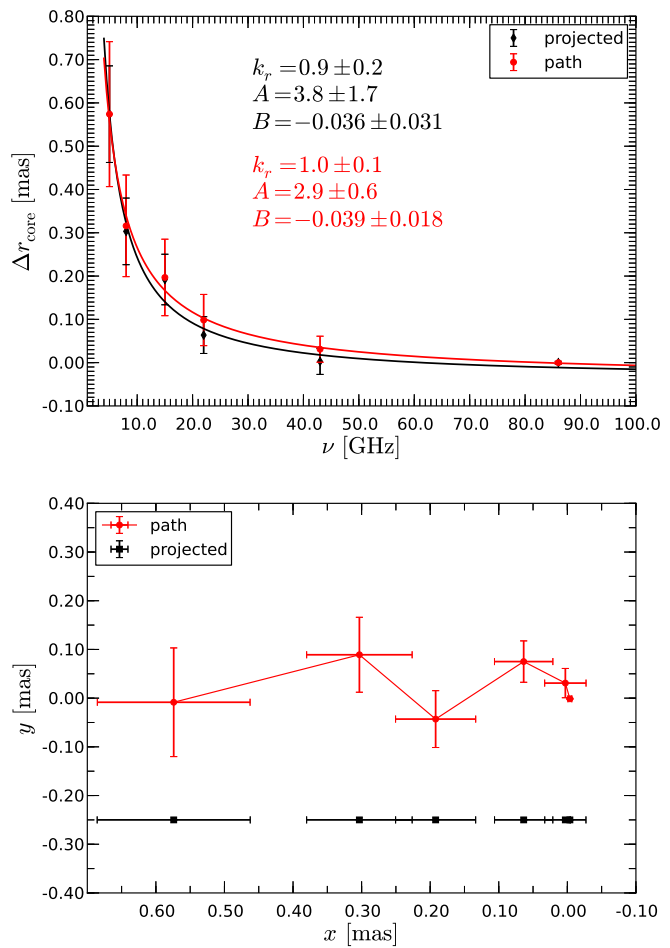}
\caption{Top: Core-shift results as a function of frequency in the jet direction for \object{CTA\,102}
at epochs 2005.4 (see Table~1). The red points correspond to the core-shifts along the 2D path and the black ones to the core-shifts projected along the average P.A. of $90^\circ$. The solid curves represent a fit to the data with the formula $\Delta r=\mathrm{A}\nu^{-1/k_r}+B$ and the values are given in the plot. Bottom: Core-shift in the sky plane. The red points correspond to the 2D path and the black ones to the projection along the average P.A. of $90^\circ$ shifted in y-direction by -0.25 mas.}
\end{figure}

 Figure~\ref{fig:totalcoreshift} displays the total core-shifts, computed by
integrating the two-dimensional path through all the intermediate frequencies. 
Our results for the first epoch show that the core does not converge to zero at our highest frequency, i.e., measurable shift between the 86\,GHz and 43\,GHz core. This result is consistent with the 86\,GHz core still corresponding to the ($\tau$=1)-surface. 


The fit of the relative position
of the core at the different frequencies with respect to the largest
($r_\mathrm{core}\propto \nu^{-1/k_r}$, Lobanov \cite{lob98})
results in a value ($k_r=1.0\pm0.1$) compatible with the expected in the
case of a conical jet in which the energies of the non-thermal particles and
the magnetic field are in equipartition, close to the minimum value
of those energies required to explain the observed radio flux. This supports the interpretation 
of the jet in \object{CTA\,102} becoming transparent to the different frequencies
as it expands.


The single-dish data from the 2006 radio flare allow us to follow the evolution of the source luminosity with time
at different frequencies. The first increase in flux density associated to the flare was detected at a turnover frequency of 222$\pm$99\,GHz
in 2005.6 (Fromm et al.\ \cite{fro11}).
Subtracting the spectrum of the source previous to the flare allows to follow the
evolution of the injected flow related to the flare. By doing this, we could
identify extra flux at frequencies larger than 100\,GHz before the main radio flare,
i.e., before the radio feature could be observed out of the
radio core at high frequencies. We can assign this extra flux to the injected
flux beyond errors and claim that it is optically thin at those frequencies before it is
at 43\,GHz or 86\,GHz. This represents independent evidence of the relation of radio-cores to optical depth in this source.
Our conclusion is that core-shift analysis performed at higher frequencies than
allowed by present techniques would result in non-convergence of the core at
a given position. We have to note that our observations indicate that a standing feature,
possibly a re-confinement shock lying 0.1\,mas away from the core at 43\,GHz.
This leaves room for the coincidence, within errors, of the core and a re-confinement
shock in other sources. 

\section{Results: black hole relative location.}
Using the results of the power law fit to the obtained core-shifts and a viewing angle of $\vartheta=2.6^\circ$ 
(Fromm et al. \cite{fro13b}), we can derive the distance of the black hole to the radio core at 86\,GHz (see Fig. 2) Following the approach of Hada et al. (\cite{had11}) the distance to the black hole is $(7.0\pm3.2)$\,pc, which is equivalent to $(8.5\pm3.9)\times10^4$ gravitational radii. In order to validate our result we compute the distance to the black hole using the projected core-shift along the average P.A. of $90^\circ$ since the calculation along the 2D path could lead to an over-estimate of the distance. The distance obtained using the projected core-shifts is $(6.4\pm5.5)$\,pc corresponding to $(7.8\pm6.7)\times10^4$ gravitational radii similar to the one using the core-shifts along the 2D path.



This distance to the black hole that we obtain fits nicely to the results obtained by Kutkin et al.\ (\cite{kut14}) and Zamaninasab et al.\ (\cite{zam13}) for the blazar 3C\,454.3 ($r_{\mathrm{core},43\,\mathrm{GHz}}\sim9\,\mathrm{pc}$). It is also comparable with the predictions for blazar and quasar jets (Marscher et al.\ \cite{mar08,mar10}) and distant radiogalaxies (Marscher et al. \cite{mar02}, Chatterjee et al. \cite{chat09,chat11}). On the contrary, it is much larger than in the case of the radio galaxy M\,87 (Hada et al.\ \cite{had11}; Marscher et al.\ \cite{mar11}).

The reason for this difference could then well be a matter of resolution and that observing M\,87 at the distance of those other radiogalaxies would bring the core to $10^{(4-5)}\,\mathrm{R_s}$. In this context, HST-1 (at 1 arcsec from the core) could be observed as the radio-core or within it of M\,87, with the particularity that it can be identified with a recollimation shock. Another relevant aspect is the viewing angle: Sources observed at small viewing angles pile up all the emission from the compact, bright regions. Therefore, if the jet brightness is high up to the radio-core, it may coincide with the last bright, projected surface, which obscures all the regions between the $\tau$=1 surface and this last surface. This effect could again be avoided by increasing resolution. 

A relevant limitation of this kind of measures is set by the uncertainties in the alignment of the highest frequency image, especially if no extended emission to align properly with lower frequencies is available. 

Despite all the difficulties associated with this calculation (see the discussion), we can claim that the separation of the radio core at tens of GHz to the black hole in CTA~102 is of the order of parsecs, implying a separation of $10^{(4-5)}\,\mathrm{R_s}$, on the basis of the values obtained for the 22~GHz (see Table~4 in Fromm et al. \cite{fro13b}), 43~GHz and 86~GHz radio-cores.  

\section{Discussion}

\subsection{Core-shifts and the 2006 flare in CTA~102}
It is difficult to derive conclusions from the influence of the passage of the component through the core region due to the large errors obtained in the calculation of the core-shifts for the affected epochs. However, a general trend that we observe by performing the same kind of analysis for the remaining epochs is that the exponent of the core-shift with frequency decreases when a perturbation crosses the core region (see Table~5 in Fromm et al. \cite{fro13b}). Actually, the crossing of the 43~GHz core by another component at the beginning of 2007 (Fromm et al. \cite{fro13a}) seems to cause the same effect as the 2006 flare (see Table~5 in Fromm et al. \cite{fro13b}). This effect can be assigned to changes in the opacity as the perturbation propagates: As the opacity increases, the high-frequency core positions are dragged downstream, and the relative core-shifts are reduced. This is reflected in the decrease of $k_r$. 

\subsection{Core-shifts and jet wobbling}

Fromm et al. (\cite{fro13b}) showed that the core-shift direction may also change in the plane of the sky. This effect could be due to \emph{jet wobbling}, which has been observed in other sources (e.g., Agudo et al. \cite{ag07}, \cite{ag12}). We would like to point out that these changes in the relative positions of the cores on the plane of the sky could result in changes in the measured projected projected core-shifts. In addition, this effect is independent from the presence of a perturbation, although a perturbation can introduce further changes. Both effects seem to be acting on the jet in CTA~102, which makes it very difficult to disentangle their relative role.

However, even the large apparent changes in direction of the core-shift vectors observed at the two first epochs presented in Fromm et al. (\cite{fro13b}, one during the steady state and another one at the first stages of the evolution of the flare) do not translate into significant changes in the value of $k_r$ obtained for both epochs ($1.0\pm0.1$ versus $0.8\pm0.3$). 

Nevertheless, the changes of the relative positions on the plane of the sky should have a negligible effect, save errors, on the calculation of the relative location of the black hole, because of the deprojection done to obtain it (Lobanov \cite{lob98}).

 \section{Summary}
 
 We present here the first 86~GHz map of CTA~102 within the series of observations around the 2006 flare. Unfortunately the other monitoring images at this frequency show poor quality. The map presented here corresponds to the first of those epochs, prior to the triggering of the flare. Making use of the simultaneous observations of the source from 5~GHz to 86~GHz and considering that the core is unaffected in this epoch by the propagation of any component, we have measured the core-shifts during what we expect to be a steady-state like situation in CTA~102. As a result, we obtain a slope for the core-shifts ($k_{\rm r}=1.0\pm0.1$) that is compatible with a conical jet in adiabatic expansion. This result allows us to compute the relative position of the 86~GHz core to the black hole, which results to be $\simeq 7\,{\rm pc}$ or $8.5\times10^4$ gravitational radii. 
 
We have discussed the possible effect of perturbations and changes on the directions of the core-shift vector and the estimates of the distance between the radio-core and the black hole. Despite the uncertainties, the same kind of calculations were performed for all epochs using the 22~GHz and 43~GHz cores as reference and provided values for the distance between them and the black-hole of the order of several parsecs in all cases, confirming our result (Fromm et al. \cite{fro13b}). Future work should include the validation of our results and possibly confirm the discussed effects on them. 

\begin{acknowledgements}
We acknowledge L. Fuhrmann for careful reading and for useful comments and suggestions to the manuscript. We thank the anonymous referee for comments and criticism that helped to improve this manuscript.
C.M.F. was supported for this research through a stipend from the International Max Planck Research School (IMPRS) for Astronomy and Astrophysics at the Universities of Bonn and Cologne.
Part of this work was supported by the COST Action MP0905 `Black Holes in a violent Universe'.
E.R. acknowledges partial support from MINECO grants AYA2009-13036-C02-02 and AYA2012-38491-C02-01 as well as Generalitat Valenciana grant PROMETEO/2009/104. M.P. acknowledges financial support from MINECO grants AYA2010-21322-C03-01, AYA2010-21097-C03-01, and CONSOLIDER2007-00050.
This work is based on observations with the VLBA, which is operated by the NRAO, a facility of the NSF under cooperative agreement by
Associated Universities Inc. This research made use of data from the MOJAVE database that is maintained by the MOJAVE team (Lister et al.\ \cite{lis09a}).
\end{acknowledgements}


\end{document}